\documentclass[11pt,twocolumn,prb,superscriptaddress,reprint]{revtex4-1}

\usepackage{graphicx,amsmath,amssymb,color}
\usepackage{tabularx, ctable}

\usepackage{ulem}
\usepackage{bm}
\usepackage{siunitx}
\usepackage{gensymb}

\usepackage[breaklinks=true,colorlinks,allcolors=blue]{hyperref}

\def\BL{\textcolor{black}}

\newcommand{\be}{\begin{eqnarray}}
\newcommand{\ee}{\end{eqnarray}}
\newcommand{\nn}{\nonumber\\}

\begin{document}

\title{Interlayer electron-hole friction 
in tunable twisted bilayer graphene semimetal}



\author{D.A. Bandurin}
\affiliation{Massachusetts Institute of Technology, Cambridge, MA 02139, USA}
\affiliation{Department of Material Science and Engineering, National University of Singapore, 117575 Singapore}

\author{A. Principi}
\affiliation{School of Physics and Astronomy, University of Manchester, Manchester M13 9PL, United Kingdom}

\author{I.Y. Phinney}
\affiliation{Massachusetts Institute of Technology, Cambridge, MA 02139, USA}
\author{T.~Taniguchi}
\affiliation{International Center for Materials Nanoarchitectonics, National Institute of Material Science, Tsukuba 305-0044, Japan}
\author{K.~Watanabe}
\affiliation{Research Center for Functional Materials, National Institute of Material Science, Tsukuba 305-0044, Japan}
 \author{P. Jarillo-Herrero}
\affiliation{Massachusetts Institute of Technology, Cambridge, MA 02139, USA}

\begin{abstract}
Charge-neutral conducting systems represent a class of materials with unusual properties governed by electron-hole (e-h) interactions. Depending on the quasiparticles statistics, band structure, and device geometry these semimetallic phases of matter can feature unconventional responses to external fields that often defy simple interpretations in terms of single-particle physics. Here we show that small-angle twisted bilayer graphene (SA-TBG) offers a highly-tunable system in which to explore interactions-limited electron conduction. By employing a dual-gated device architecture we tune our devices from a non-degenerate charge-neutral Dirac fluid to a compensated two-component e-h Fermi liquid where spatially separated electrons and holes experience strong mutual friction. This friction is revealed through the $T^2$ resistivity that accurately follows the e-h drag theory we develop. Our results provide a textbook illustration of a smooth transition between different interaction-limited transport regimes and clarify the conduction mechanisms in charge-neutral SA-TBG.

\end{abstract}

\maketitle

Low-dimensional electron-hole (e-h) systems have recently emerged as an important platform in which to explore many-body quantum phenomena. In such systems, strong Coulomb interaction among electrons and holes can give rise to a plethora of exotic quantum phases whose inventory encompasses superfluids~\cite{Eisenstein2004,AllanSuperfluid}, correlated density wave states~\cite{RickhausEH,Zarenia2017}, excitonic insulators~\cite{exciton_insulator,zhang2021correlated}, and Wigner crystals~\cite{Rakhmanov1978,Zarenia2017}, to name a few. Particularly interesting interacting e-h mixtures are hosted by graphene and its bilayer. Graphene-based devices enabled the discovery of novel non-trivial effects governed by e-h interactions: from the Wiedemann-Franz law violation~\cite{Crossno} and the anomalous Coulomb drag~\cite{LPDragGraphene,NegativeCoulomb,Song2013,GiantDrag,TitovTheory,LevitovDrag} to the quantum critical conductivity~\cite{QuantumCritFritz,Schmalian,Gallagher} and giant thermal diffusivity~\cite{heatspread}. Central in these effects is the dominance of momentum-conserving e-h collisions over other momentum-relaxing scattering processes brought upon by graphene's weak electron-phonon coupling and low disorder~\cite{Lucas}. As a result, the behavior of graphene's e-h plasma at elevated temperatures $T$, often referred to as Dirac fluid, resembles that of interacting relativistic fluids governed by the laws of (relativistic) hydrodynamics~\cite{Crossno,Lucas,SvintsovEh,Narozhny,ShaffiqueBLG}. Since hydrodynamics offers a natural framework by which to probe the long-wavelength behavior of strongly-interacting fluids, experiments on model platforms, such as graphene, can give insights for observations in more exotic quantum phases of matter~\cite{ColdAtoms,Quark}, substantiating the interest in the field. 

So far, the hydrodynamic behavior of interacting e-h plasmas in mono- and bilayer graphene (MLG and BLG respectively) was explored deep in the non-degenerate limit ($E_\mathrm{F}\ll k_\mathrm{B} T$, where $E_\mathrm{F}$ is the Fermi energy, $k_\mathrm{B}$ is the Boltzmann constant) and relied on thermal~\cite{Crossno,Morpurgo,JHone_BLG}, light-~\cite{heatspread} or current-driven~\cite{berdyugin2021outofequilibrium} excitation of e-h pairs. The ambipolar hydrodynamics in the degenerate regime ($E_\mathrm{F}\gg k_\mathrm{B} T$) as well as its genesis from the Boltzmann phase have at present remained inaccessible. This inaccessibility stems from the fact that the conduction and valence band extrema in MLG and BLG coincide in momentum space and thus the e-h system can only be realized through the smearing of the charge neutrality point (NP)$;$ adding more carriers into the system converts the neutral Dirac fluid into a unipolar Fermi liquid (FL)~\cite{Lucas}. In this work, we introduce biased SA-TBG as a convenient system in which to explore a smooth crossover between the Dirac fluid regime and the regime of degenerate e-h FL. In the latter case, we demonstrate that frequent momentum-conserving (yet velocity-relaxing) e-h collisions are the limiting factor for the SA-TBG conductivity.

\begin{figure}[ht!]
  \centering\includegraphics[width=1\linewidth]{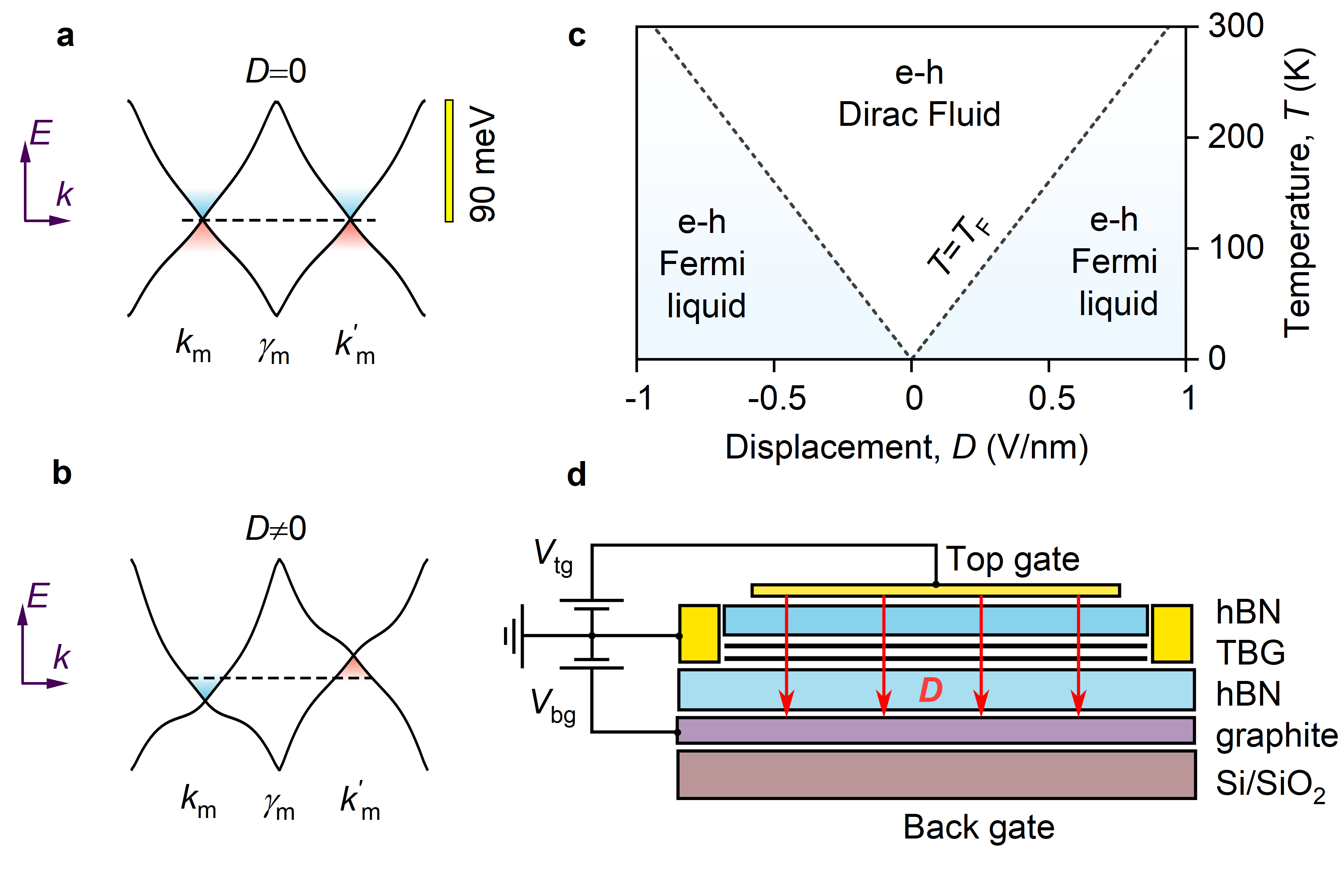}
    \caption{\textbf{Biased SA-TBG.}
    \textbf{a-b,} Calculated single-particle band structure for 1.65$\degree$ SA-TBG~\cite{Kaxiras1,Kaxiras2}. At low-energies, two Dirac cones are formed in the vicinity of the $k_\mathrm{m}$ and $k'_\mathrm{m}$ points (a); when $D\neq 0$, the cones are shifted with respect to each other (b). The horizontal dashed lines represent the Fermi level in the neutral SA-TBG. \textbf{c,} Phase diagram for the charge-neutral e-h mixture in SA-TBG mapped onto a $T-D$ plane. Dashed lines: the dependence of $T_\mathrm{F}$ in each minivalley on $D$ for $n=0$. \textbf{d,} Schematic of the dual-gated encapsulated SA-TBG device. }
	\label{Fig1}
\end{figure}

\begin{figure*}[ht!]
  \centering\includegraphics[width=0.98\linewidth]{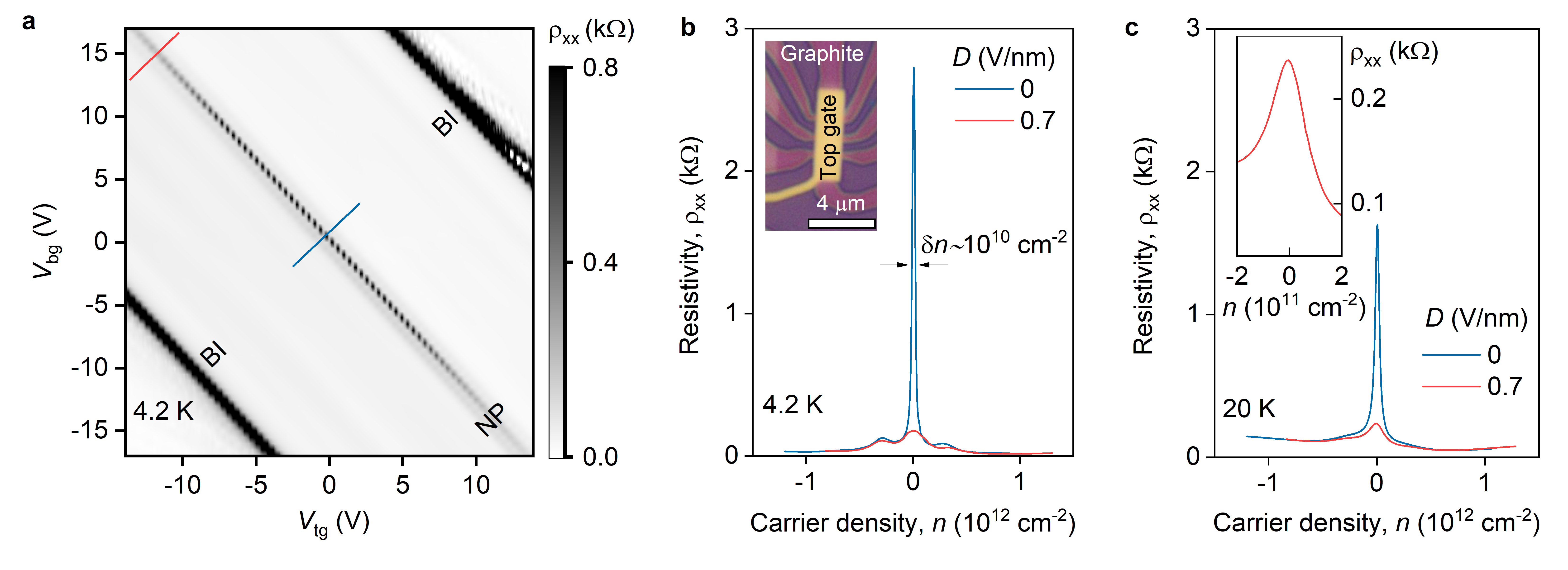}
    \caption{\textbf{Effect of displacement on the transport properties of the SA-TBG. } \textbf{a,} $\rho_\mathrm{xx}$ as a function of $V_\mathrm{bg}$ and $V_\mathrm{tg}$ measured in the 1.65$\degree$ SA-TBG device. Blue and red lines correspond to the $(V_\mathrm{tg},V_\mathrm{bg})$ points where $D=0$ and $D=0.7~$V/nm respectively. \textbf{b,} $\rho_\mathrm{xx}(n)$ traces for $D=0~$ and $D=0.7~$V/nm measured at $T=4.2~$K. Inset: Optical photograph of an encapsulated SA-TBG device.  \textbf{c,} Same as (b) but for $T=20~$K. Inset: zoomed-in region of the NP vicinity for $D=0.7~$V/nm. }
    \label{Fig2}
\end{figure*}

We start by exploring the single-particle band structure of SA-TBG which is folded within a reduced Brillouin zone (BZ)~\cite{bistritzer2011} due to superlattice periodicity (Fig.~\ref{Fig1}a-b). At small energies, it resembles that of MLG but is characterized by a decreased Fermi velocity $v_\mathrm{F}$. Like the BZ of MLG, the reduced BZ of SA-TBG is hexagonal and comprises two minivalleys located at the $k_\mathrm{m}$ and $k'_\mathrm{m}$ high symmetry points. These coincide with the $K$ points of the two decoupled graphene sheets~\cite{bistritzer2011}. A prominent feature of the SA-TBG is that, away from the magic angle \BL{($\theta\gtrsim1.3\degree$)}, one can selectively populate its minivalleys with charge carriers of opposite types using a perpendicular displacement field, $D$, (Fig.~\ref{Fig1}b)~\cite{SanchezPRL,Slizovskiy,Rickhaus_minivalley,berdyugin2020,IzzzzzyPRL,RickhausEH}. Electrostatic  calculations~\cite{Slizovskiy} for $D=1~$V/nm, reveal that such a strong $D$, readily achievable in experiments, can result in the formation of relatively large electron and hole Fermi surfaces in the $k_\mathrm{m}$ and $k'_\mathrm{m}$ minivalleys, respectively. Quantitatively, in each minivalley, the Fermi temperature, $T_\mathrm{F}$, exceeds room $T$, as in normal
FLs (Fig.~\ref{Fig1}c dashed line). On the contrary,  charge-neutral SA-TBG at $D=0$ is half-filled up to the Dirac point where the Fermi surfaces shrink to two points and where the Dirac fluid emerges at elevated $T$~\cite{Crossno,Lucas}. This tunability enables the exploration of e-h plasma at the crossover between the Dirac fluid and FL regimes in standard transport experiments as we schematically illustrate on the $D-T$  diagram in Fig~\ref{Fig1}c.

To probe such a crossover, we fabricated a dual-gated multi-terminal Hall bar made out of $\theta\approx1.65\degree$ SA-TBG encapsulated between two relatively thin ($<100$~nm thick) slabs of hexagonal boron nitride (hBN). \BL{At this angle, the SA-TBG is characterized by enhanced interaction strength and a reduced $v_F$, but is far enough from the magic angle ($1.1\degree$) that it allows for appreciable interlayer polarization~\cite{Rickhaus_minivalley,RickhausEH}}. The device was produced by a combination of tear-and-stack~\cite{TutucNL2016,Tutuc_TBG,YuanPRL} and hot release~\cite{Hot-transfer_NComm} methods, and had a width of $2$ $\mu$m (Inset of Fig.~\ref{Fig2}b) (Supplementary Section 1). The dual-gated configuration (Fig.~\ref{Fig1}d) allowed us to control the interlayer displacement $D/\epsilon_\mathrm{0}=(C_\mathrm{bg}V_\mathrm{bg}-C_\mathrm{tg}V_\mathrm{tg})/2$, and the total externally-induced carrier density, $n=(C_\mathrm{bg}V_\mathrm{bg}+C_\mathrm{tg}V_\mathrm{tg})/e$, where $C_\mathrm{tg,bg}$ are the top and bottom gate capacitance per unit area, $\epsilon_\mathrm{0}$ is the dielectric permittivity of vacuum, and $e$ is the electron charge. 

Figure~\ref{Fig2}a shows an example of the longitudinal resistivity, $\rho_{xx}$, dependence on $V_\mathrm{bg}$ and $V_\mathrm{tg}$ in a form of 2D map, measured in our SA-TBG and reveals its characteristic behavior. Namely, the map consists of three diagonal lines: central - that denotes the global neutrality, and two side diagonals, labeled as BI, that reflect the full filling of the first miniband where the single-particle band insulator emerges~\cite{Tutuc_TBG,YuanPRL,SmetTBG}. The BI lines allow for an accurate determination of the twist angle~\cite{Tutuc_TBG,YuanPRL,SmetTBG}. \BL{Below, we will only focus on the region in the vicinity of the global neutrality and away from the van Hove singularity.}

Figure~\ref{Fig2}b shows the $\rho_\mathrm{xx}(n)$ dependence of our SA-TBG device measured at $D=0$ and $T=4.2~$K (the curve is measured along the blue trace in the map from Fig.~\ref{Fig2}a). At $D=0$, 
$\rho_\mathrm{xx}(n)$ exhibits a sharp peak and reaches 2.7~k$\Omega$ at $n=0$, a standard behavior for SA-TBG devices. The peak width is only $\delta n \simeq \times10^{10}$~cm$^{-2}$ that indicates low charge inhomogeneity provided by the graphite gate~\cite{Zibrov}. Upon doping, $\rho_\mathrm{xx}(n)$ rapidly decreases and already at $10^{12}~$cm$^{-2}$ drops to 30$~\Omega$ which translates to the $1.7~\mu$m mean free path, obtained from the standard Drude model. At liquid helium $T$, we also observed negative transfer resistance measured in the bend geometry (Supplementary Information 2), an indicative of the micrometre-scale ballistic transport~\cite{Beenakker,Mayorov2011}. These observations highlight an exceptional quality of our encapsulated SA-TBG device critical for further exploration of interaction-dominated transport at elevated $T$ as we now proceed to discuss. 

\begin{figure*}[ht!]
	\centering\includegraphics[width=0.85\linewidth]{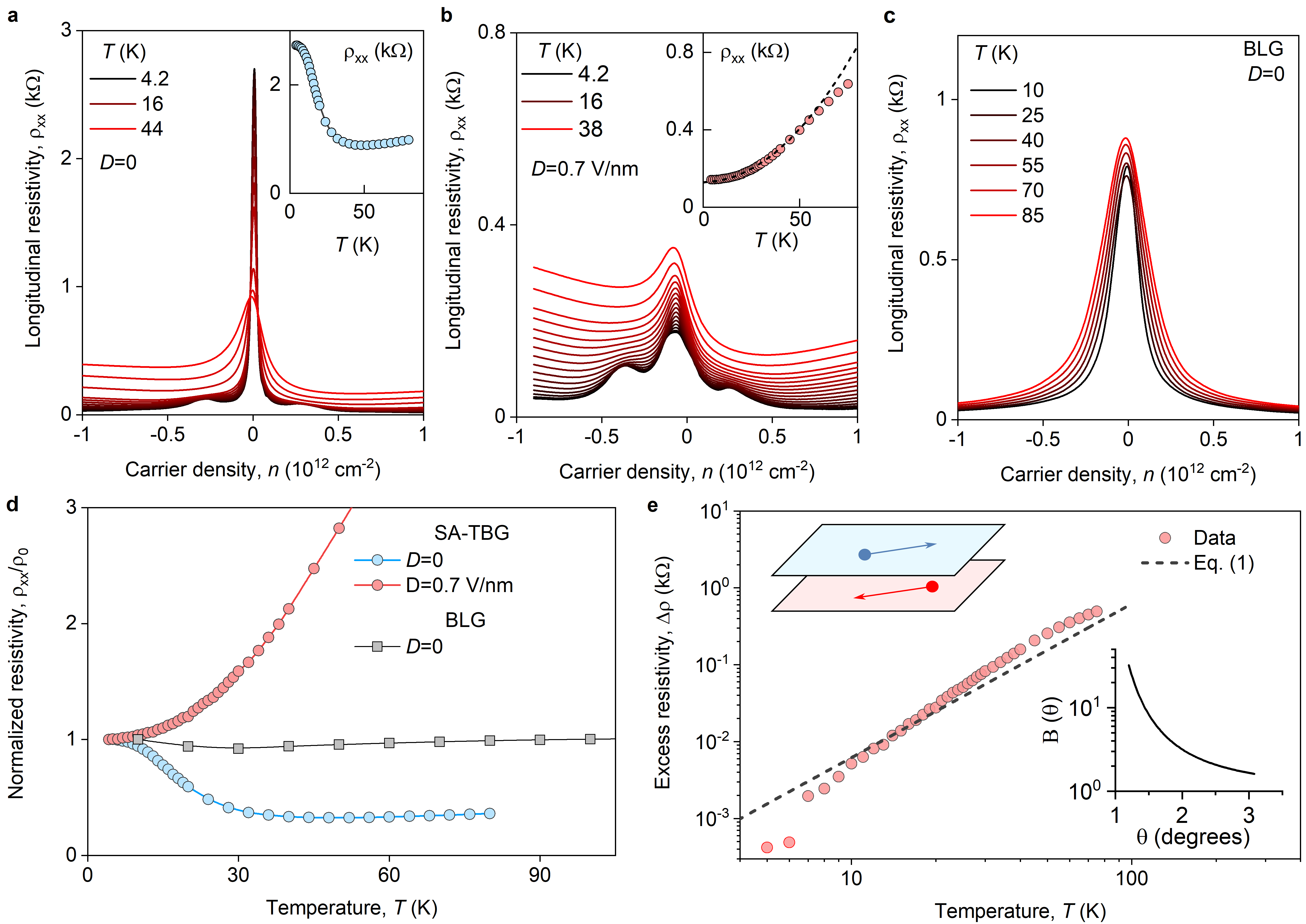}
	\caption{\textbf{Temperature dependence of the SA-TBG resistivity} \textbf{a,} $\rho_\mathrm{xx}(n)$ for different $T$ for the case of $D=0$. Inset: $\rho_\mathrm{xx}(T)$ at the NP and $D=0$. \textbf{b,} Same as (a) but for $D=0.7~$V/nm. Inset: $\rho_\mathrm{xx}(T)$ at the compensation point ($n=0$) and $D=0.7~$V/nm. Dashed line: guide for the eye that represents the $a+b T^2$ dependence. \textbf{c,} $\rho_\mathrm{xx}(n)$ for BLG at $D=0$.  \textbf{d,} Resistivity as a function of $T$ for the charge-neutral SA-TBG at $D=0$ (blue) and $D=0.7~$V/nm (red) and for BLG at $D=0$ (grey). The data is normalized to the lowest-$T$ value of $\rho_\mathrm{xx}(n)$: 4.2~K for SA-TBG and 10~K for BLG. \textbf{e,} $\Delta \rho=\rho_\mathrm{xx}(T)-\rho_\mathrm{xx}(4.2~K)$ as a function of $T$ measured at $D=0.7~$V/nm and $n=0$ (symbols). Note, $\Delta \rho(T)$ exhibits somewhat faster $T-$dependence at $T<15~$K. This apparent behavior is spurious and is related to the subtraction operation of the $\rho_0=\rho_\mathrm{xx}(4.2~K)$ from the experimental dataset rather than $\rho_\mathrm{xx}$ at $T\rightarrow0$. Solid line: theoretical dependence, eq. (1). Upper left inset: schematic illustration of the interlayer e-h friction in SA-TBG at finite $D$. Lower right inset: Prefactor ${\cal B}$ as a function of twist angle, $\theta$. }
	\label{Fig3}
\end{figure*}

With the application of $D$, the transport properties of neutral SA-TBG change drastically (Fig.~\ref{Fig2}b, red curve). $\rho_\mathrm{xx}$ at the NP drops by more than an order of magnitude and becomes comparable to that of doped SA-TBG (cf. $\rho_\mathrm{xx}$ at $10^{12}~$cm$^{-2}$). This qualitative behavior remains unchanged upon increasing $T$ (Fig.~\ref{Fig2}c). Namely, at $T=20~K$ the NP resistivities measured at zero and finite $D$ differ by more than an order of magnitude. The drop of $\rho_\mathrm{xx}$ with increasing $D$ signals parallel conduction of two minivalleys when each of them is doped away from their NPs.

We further studied the temperature dependence of our sample's resistivity.  Figures~\ref{Fig3}a-b shows $\rho_\mathrm{xx}(n)$ dependencies for varying $T$ for the case of zero (a) and finite (b) $D$ respectively. Away from the NP ($n=0$), $\rho_\mathrm{xx}$ grows with increasing $T$ for both $D$ values, indicating characteristic behavior of doped graphene sheets. 
On the contrary, at the NP, $\rho_\mathrm{xx}$ exhibits a very different behavior for the two cases. Namely, at $D=0$, $\rho_\mathrm{xx}$ drops rapidly when $T$ is raised from 4.2 to 40 K (inset of Fig.~\ref{Fig3}a), whereas at $D=0.7~$V/nm, $\rho_\mathrm{xx}$ shows a clear metallic trend: the resistivity increases with increasing $T$ (inset of Fig.~\ref{Fig3}b).

It is now instructive to normalize all measured $\rho_\mathrm{xx}(T)$ dependencies to their lowest $T$ value in order to compare the functional forms of the  $T-$dependencies in different cases. At $T=40$~K, the zero-$D$ resistivity of the SA-TBG device is less than a half of its $4.2~$K value; further increase of $T$ leads to a very slow ascending trend of $\rho_\mathrm{xx}(T)$. At the same $T$ and $D=0.7~$V/nm, $\rho_\mathrm{xx}$ experiences more than two times increase and keeps growing with increasing $T$ following approximately an $a+b T^2$ dependence, where $a$ and $b$ are constants (dashed black line in the inset of Fig.~\ref{Fig3}b). To compare, we have also measured the resistivity of a BLG device of comparable quality as a function of $n$ and $T$ (Fig.~\ref{Fig3}c). At the NP, $\rho_\mathrm{xx}$ is practically unaffected by the $T$ variation (Fig.~\ref{Fig3}c) over the entire range of $T$ in our experiments.

The above observations clearly point to the difference in the conductivity mechanisms of these three bilayer systems at their NPs. The weak insulating behavior of charge-neutral SA-TBG at zero $D$ resembles that of MLG: the resistivity drops as a result of the thermal activation of electrons and holes~\cite{Crossno}. A further increase of $T$ leads to the enhanced scattering between electron and hole non-degenerate sub-systems hosted by SA-TBG leading to an increase of the resistivity. In contrast, the flat $T-$dependence of the BLG has been recently attributed to the perfect balance between the amount of thermally activated e-h pairs facilitating conductivity, and the e-h scattering that impedes the electrical current\cite{JHone_BLG,ShaffiqueBLG,BLG-toy}. The peculiar $T^2$ growth of the resistivity in compensated SA-TBG at finite $D$ has not been observed previously. Below we show that this effect stems from the e-h friction~\cite{Kvon2,Muller_nanolett_2021} in this degenerate ambipolar system.

To demonstrate this, we solve the steady-state Boltzmann equation for e-h hole mixture in SA-TBG; the details are given in Supplementary Information.
In the limit of temperatures much smaller than $T_\mathrm{F}$,
the resistivity due to e-h scattering reads
%
%
%
\be
\rho_{\rm D} &\simeq&
\frac{8\pi \alpha_{\rm ee}^2 g({\bar q}_{\rm TF}) }{3 n e^2 v_{\rm F}^2 \hbar }
 (k_{\rm B} T)^2
~.
\ee
where 
$n$ is the particle density in each minivalley, 
$g({\bar q}_{\rm TF}) = 3({\bar q}_{\rm TF} - 1)+ (4 - 3 {\bar q}_{\rm TF}^2) {\rm arccoth}(1+{\bar q}_{\rm TF})$ and ${\bar q}_{\rm TF} = N_{\rm f}\alpha_{\rm ee}$ is the Thomas-Fermi screening wavevector in units of the Fermi wavevector. Here, $\alpha_{\rm ee} = e^2/(2\pi \epsilon_0 (\epsilon_r+1) \hbar v_{\rm F})$ is the effective fine-structure constant of Dirac fermions, $\epsilon_r$ is a dielectric constant accounting for screening due to far bands and external dielectrics, $N_{\rm f}$ is the number of flavors, and $\hbar$ is the reduced Planck constant. Hereafter we set $\epsilon_r = 3.9$, as for graphene deposited on hBN. The total resistivity is then $\rho = \rho_0 + \rho_{\rm D}$, where $\rho_0$ is the zero-temperature resistivity due to momentum-non-conserving scattering processes. We also note that, as the minivalleys are predominantly formed from the energy bands of different graphene sheets, electrons and holes reside in the upper or lower graphene layers depending on the $D$ direction~\cite{Slizovskiy, Rickhaus_minivalley,RickhausEH}, and thus $\rho_\mathrm{D}$ can be interpreted as the resistivity due to the interlayer e-h friction.

In Fig.~\ref{Fig3}d we compare the results of our calculations with $\rho_\mathrm{xx}(T)$ found experimentally. To this end, we plot the experimentally found resistivity excess, $\Delta \rho=\rho_\mathrm{xx}(T)-\rho_\mathrm{xx}(4.2~K)$, and theoretically obtained $\rho_{\rm D}(T)$. For the latter,  we used an electrostatic model that accounts for screening effects to calculate the Fermi energy in each minivalley~\cite{Slizovskiy}, as well as the experimentally determined twist angle. Using that, for $\theta = 1.65^\circ$, $v_{\rm F} \simeq 5\times 10^5~{\rm m/s}$ (as determined from the continuum model of SA-TBG~\cite{bistritzer2011,Kaxiras1,Kaxiras2}), \BL{for $D = 0.7~{\rm V/m}$ we estimate the carrier density $n=1.3\times 10^{15}~{\rm m}^{-2}$}. Experimental data follows closely the expected ${\cal B}T^2$ dependence with ${\cal B}\simeq0.062~\Omega$/K$^2$ with some tendency to sub-quadratic dependence at higher $T$ (inset of Fig.~\ref{Fig3}b). This deviation from the $T^2$ scaling can be attributed to the thermal smearing of the distribution function that leads to the exit of the SA-TBG e-h system from the degenerate state. Indeed, at $n=1.3\times 10^{11}~{\rm  cm}^{-2}$, the Fermi temperature of the $1.65\degree$~SA-TBG is of the order of $220~$K.

Next, we analyze $\rho_{\rm D}(T)$ dependencies expected for other $\theta$. We find that, at fixed carrier density, the resistivity due to e-h scattering depends on $\theta$ only through its dependence on the electron Fermi velocity $v_{\rm F}$. The latter controls the values of both the Fermi energy $\varepsilon_{\rm F}$ and effective fine-structure constant $\alpha_{\rm ee}$. The relationship between the Fermi velocity and twist angle can be obtained from a continuum model of SA-TBG~\cite{bistritzer2011}. In the inset of Fig.~\ref{Fig3} we plot the ratio ${\cal B}(v_{\rm F})/{\cal B}(v_{\rm F}^{\rm g})$ for a carrier density $n=4\times 10^{14}~{\rm m}^{-2}$ as a function of $\theta$. Here, $v_{\rm F}^{\rm g}$ is the Fermi velocity of MLG, while ${\cal B}(v_{\rm F})$ is defined from $\rho_{\rm D} = {\cal B}(v_{\rm F}) T^2$. At $\theta>3\degree$ the e-h drag would result in a 10 times smaller prefactor of the $T^2-$ resistivity with respect to that observed in the present experiment on SA-TBG. Indeed, at large $\theta$, the layers are fully decoupled and $v_{\rm F}~\approx v_{\rm F}^{\rm g}$. On the contrary, close to the magic angle, the layers are hybridized so that control of individual minivallyes via electrostatic means cannot be realized.

It would be instructive to put our observations in the context of electron transport in semimetals. 
Depending on quasiparticle statistics, band structure details, device geometry and interaction strength, seemingly alike semimetallic e-h systems can display very different physical properties and regimes of transport. For example, in charge-neutral MLG, frequent collisions between thermally activated electrons and holes impede electrical currents while leaving thermal ones untouched, causing a staggering breakdown of the Wiedeman-Franz law. In this system the Lorentz ratio, i.e. the ratio between the thermal conductivity and its electrical counterpart, is found to be greatly enhanced~\cite{Crossno}. 
On the contrary, in degenerate compensated semimetals such as WP$_2$ or Sb the Lorentz ratio has been found to be suppressed~\cite{Gooth}. Despite their semimetallic nature, which would imply violations of the Wiedeman-Franz law akin to those observed in graphene~\cite{Ale2020}, the behavior of these materials closely resembles that of conventional unipolar systems~\cite{Principi_prl_2015}, where carriers of a single type transport both charge and heat. All these seemingly contradictory observations have stimulated a debate over the effect of quasiparticle statistics, band structure and many-body interactions on the thermal and electrical properties of these charge-neutral material platforms~\cite{MaslovWF, Ale2020, Alekseev}. A definitive resolution of these long lasting puzzles is made especially difficult by the fact that completely different  behaviors are observed in different systems and regimes, and therefore a thorough comparison between them becomes challenging. The behavior of SA-TBG observed in this work  thus makes it a highly-tunable platform for the exploration of different semimetallic regimes on an equal footing, allowing for a gradual transition between them. 

To conclude, we have shown that SA-TBG offers a highly-tunable semimetalic system in which to explore physics at the crossover between the charge-neutral Dirac fluid and compensated e-h FL. In the latter case we found strong Coulomb friction between spatially separated electron and hole subsystems that resulted in the $T^2-$ growth of the resistivity. Finally, we have developed a theory for e-h scattering in SA-TBG and found that its predictions are close to the experimental observations.  It would be further interesting to explore transport and thermal properties of e-h FLs in other polarizable layered systems with heavier charge carriers such as twisted double bilayer graphene~\cite{RickhausEH} or twisted transition metal dichalcogenides~\cite{zhang2021correlated} as well as to explore collective modes in such e-h mixtures~\cite{Andreeva,Euler,Fogler}.

\vspace{1em}

\noindent\rule{6cm}{0.4pt}

*Correspondence to: dab@nus.edu.sg, alepr85@gmail.com and pjarillo@mit.edu.




\section*{Acknowledgements}
This work was supported by AFOSR grant FA9550-21-1-0319 and the Gordon and Betty Moore Foundation’s EPiQS Initiative through Grant GBMF9463 to P.J.-H. D.A.B. acknowledges the support from MIT Pappalardo Fellowship. I.Y.P acknowledges support from the MIT undergraduate research opportunities program and the Johnson and Johnson research scholars program. K.W. and T.T. acknowledge support from JSPS KAKENHI (Grant Numbers 19H05790, 20H00354 and 21H05233). A.P. acknowledges support from the European Commission under the EU Horizon 2020 MSCA-RISE-2019 programme (project 873028 HYDROTRONICS) and from the Leverhulme Trust under the grant RPG-2019-363. The authors thank Clement Collignon, Alexey Berdyugin and Dmitry Svintsov for productive discussions.

\section*{Data availability}
All data supporting this study and its findings are available within the article and its Supplementary Information or from the corresponding authors upon reasonable request.

\section*{Author contributions}
D.A.B. and A.P. conceived and designed the study. D.A.B. and I.Y.P. fabricated and measured the devices. T.T. and K.W. grew high-quality hBN crystals. A.P. developed the theory. P.J.H. supervised the project.

\section*{Competing interests}
The authors declare no competing interests.


%

\newpage
\setcounter{figure}{0}
\renewcommand{\thesection}{}
\renewcommand{\thesubsection}{S\arabic{subsection}}
\renewcommand{\theequation} {S\arabic{equation}}
\renewcommand{\thefigure} {S\arabic{figure}}
\renewcommand{\thetable} {S\arabic{table}}

\begin{widetext}

\section*{\textbf{Supplementary Information}}
\subsection{Device fabrication.}

Our device consisted of hBN-encapsulated twisted bilayer graphene, which we fabricated using a combination of cut-and-stack~\cite{Tutuc_TBG, lasercut} and hot release~\cite{Hot-transfer_NComm} methods. Monolayer graphene, few-layer graphite, and 30-80\,nm-thick hBN crystals were mechanically exfoliated on a Si/SiO$_2$ substrate, and sizable, uniform flakes were selected using optical microscopy. Then, using a homemade transfer system with $\mu$m-accuracy and a polycarbonate (PC) membrane stretched over a small (8\,mm$\times$8\,mm$\times$4\,mm) polydimethylsiloxane (PDMS) polymer block on a glass slide, we assembled hBN and graphite stacks on a Si/SiO$_2$ wafer. We first picked up hBN crystal at 50-70\,\degree C. Then, the graphite was picked up at room temperature, and then the entire stack was ``ironed'' and then released on a clean Si/SiO$_2$ wafer at high temperatures ($160~-~170$\,\degree C). After removing the polymer membrane, we annealed the hBN and graphite stack at 350\,\degree C for 3 hours in argon/hydrogen atmosphere. We then assembled the hBN and twisted bilayer graphene stack using a ``cut-and-stack'' method described previously~\cite{Tutuc_TBG,YuanPRL}. After picking up the top hBN and twisted graphene, we ``ironed'' the entire stack at room temperature. The three-layer stack was then released onto the previously fabricated and cleaned bottom hBN and graphite gate at roughly 160\,C. After this point, we avoided heating the stack to reduce the possibility of twist angle relaxation. The final stack was inspected using dark-field microscopy and atomic force microscopy (AFM), and bubble- and blister-free areas were selected to use for Hall bars.

To fabricate the devices, we covered the heterostructures by a protective polymethyl-methacrylate (PMMA) resist and used electron beam lithography (EBL) to define contact regions. We then performed a mild O$_2$ plasma cleaning before using reactive ion etching (RIE) with a plasma generated from  CHF$_3$ and O$_2$ gases to selectively etch away the hBN in the parts of the heterostructure unprotected by the lithographic mask~\cite{MosheNatPhys}. 3\,nm chromium and 50-70\,nm gold was then evaporated into the contact regions via thermal evaporation at high vacuum. We repeat the same EBL and thermal evaporation procedures to define a metallic top gate (3\,nm chromium and 30-40\,nm gold). Finally, we repeat the same EBL and RIE procedures to define the final Hall bar geometry, using, in this case, a plasma generated by Ar, O$_2$ and CHF$_3$ gases.

\newpage
\subsection{Signatures of ballistic transport in SA-TBG.}

One of the standard ways to probe the presence of ballistic transport in mesoscopic devices is to measure the transfer resistance in the bend geometry illustrated in Fig. S1~\cite{Beenakker,Mayorov}. In this geometry, an electrical current, $I_\mathrm{1-2}$, is passed between contacts 1 and 2 and the voltage drop, $I_\mathrm{3-4}$, is measured between contacts 3 and 4. In the case of diffusive transport, this configuration yields a positive signal, $R_\mathrm{3-4,1-2}=V_\mathrm{3-4}/I_\mathrm{1-2}$ because charge carriers flow along electric field lines.  The bend geometry is then topologically identical to the conventional 4-point configuration used for resistivity measurements. On the contrary, if the charge carriers experience ballistic motion on a scale of the device width then nothing prevents them from reaching the opposite device boundary producing negative $R_\mathrm{3-4,1-2}$. The negative sign of the transfer resistance measured in the bend geometry is usually considered as a benchmark of ballistic transport regime. 

\begin{figure*}[ht!]
	\centering\includegraphics[width=0.4
	\linewidth]{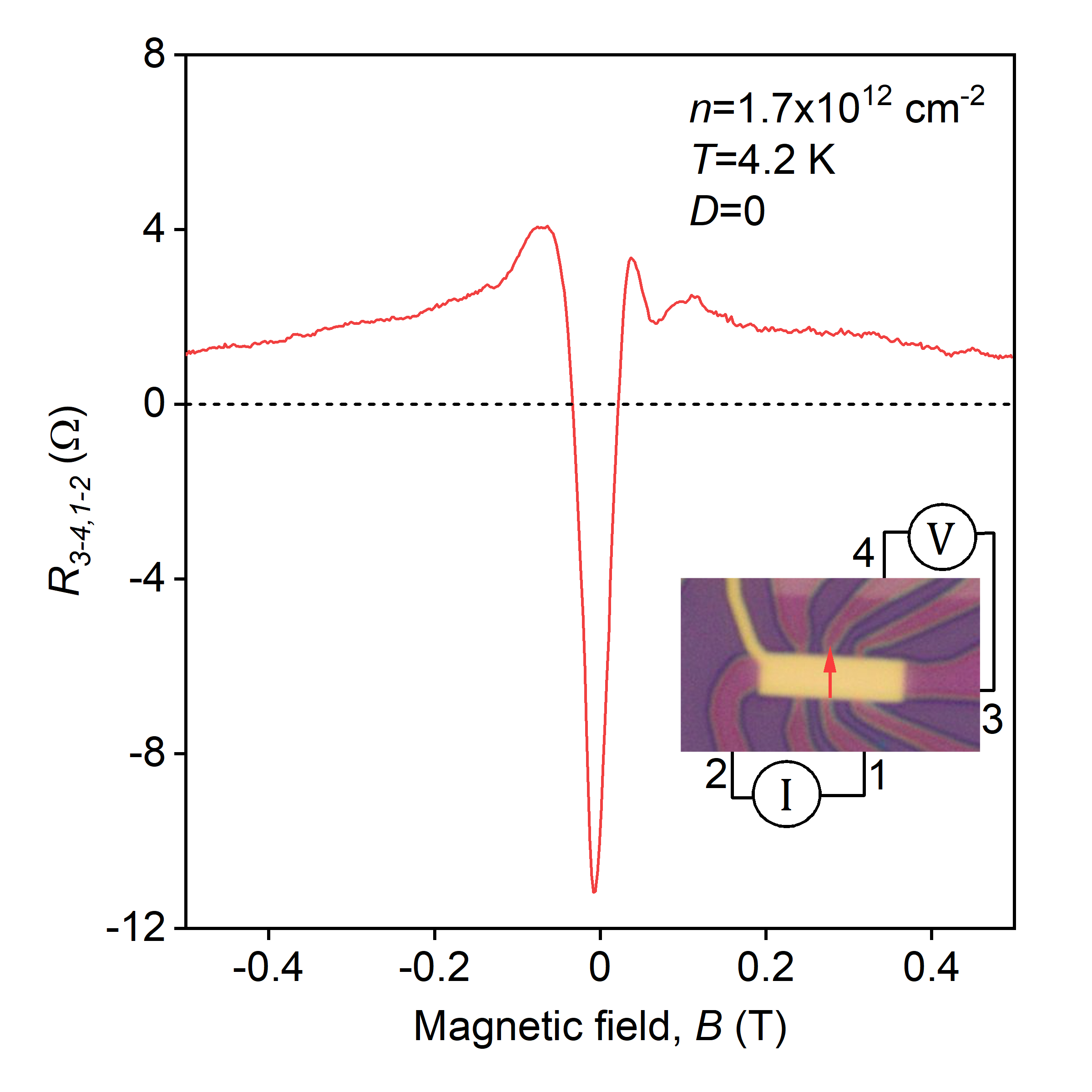}
	\caption{\textbf{Micrometer-scale ballistic transport in encapsulated SA-TBG.} $R_\mathrm{3-4,1-2}$ as a function of magnetic field measured in the geometry, shown in the inset, for given $n,T$ and $D$. }
	\label{FigS2}
\end{figure*}

We have performed such measurements in our sample and found that at liquid helium $T$, $R_{3-5,1-2}$ is negative over the wide range of $n$. Furthermore, the application of a magnetic field, causes the sign change of the measured signal, as the charge carriers are deflected from the straight trajectories. These observations highlight a high quality of our encapsulated sample, that is crucial for studies of interaction-dominated transport described in the main text. 

\newpage
\subsection{Theoretical calculations of the interaction-dominated resistivity in SA-TBG.}
To derive the resistivity due to electron-electron interactions, we start from the Boltzmann equation for the space- and time-dependent fermion occupation function $f_{{\bm k},\lambda}({\bm r},t)$,
\begin{eqnarray} \label{eq:boltzmann_def}
\partial_t f_{{\bm k},\lambda}({\bm r},t) + {\bm v}_{{\bm k},\lambda}\cdot {\bm \nabla}_{\bm r} f_{{\bm k},\lambda}({\bm r},t) - e {\bm E}\cdot {\bm \nabla}_{\bm k} f_{{\bm k},\lambda}({\bm r},t) = {\cal I}_{\rm ee}[f_{{\bm k},\lambda}]
~,
\end{eqnarray}
where ${\bm v}_{{\bm k},\lambda} = {\bm \nabla}_{\bm k}\varepsilon_{{\bm k},\lambda}$ is the velocity of a particle of Bloch wavevector ${\bm k}$ characterized by (band, spin, valley and layer) quantum numbers $\lambda$, $-e$ is the electron charge, and ${\bm E}$ is the (uniform and time-independent) external electric field. Finally, ${\cal I}_{\rm ee}[f_{{\bm k},\lambda}]$ is the collision integral of electron-electron interactions:
\begin{eqnarray} \label{eq:I_ee_def}
{\cal I}_{\rm ee}[f_{{\bm k},\lambda}] &=& \frac{1}{{\cal A}^3} \sum_{\substack{{\bm k}_2,\lambda_2}} \sum_{\substack{{\bm k}_3,\lambda_3\\{\bm k}_4,\lambda_4}} W_{\rm ee}({\bm k}_1, \lambda_1; {\bm k}_2, \lambda_2; {\bm k}_3, \lambda_3; {\bm k}_4, \lambda_4) \delta({\bm k}_1+{\bm k}_2-{\bm k}_3-{\bm k}_4) \delta(\varepsilon_{{\bm k}_1,\lambda_1}+\varepsilon_{{\bm k}_2,\lambda_2}-\varepsilon_{{\bm k}_3,\lambda_3}-\varepsilon_{{\bm k}_4,\lambda_4})
\nonumber\\
&\times&
\big[ f_{{\bm k}_1,\lambda_1} f_{{\bm k}_2,\lambda_2} (1-f_{{\bm k}_3,\lambda_3})(1-f_{{\bm k}_4,\lambda_4}) - (1-f_{{\bm k}_1,\lambda_1})(1-f_{{\bm k}_2,\lambda_2}) f_{{\bm k}_3,\lambda_3} f_{{\bm k}_4,\lambda_4} \big]
~.
\end{eqnarray}
Within the Fermi-golden rule,
\begin{eqnarray} \label{eq:Wee_def}
W_{\rm ee}({\bm k}_1, \lambda_1; {\bm k}_2, \lambda_2; {\bm k}_3, \lambda_3; {\bm k}_4, \lambda_4) &=& \frac{2\pi}{\hbar} |V_{\rm ee}({{\bm k_1}-{\bm k_3}},\varepsilon_{{\bm k}_1,\lambda_1} - \varepsilon_{{\bm k}_3,\lambda_3})|^2 
{\cal D}({{\bm k}_1,\lambda_1}; {\bm k}_3,\lambda_3) {\cal D}({{\bm k}_2,\lambda_2}; {\bm k}_4,\lambda_4)
~,
\nonumber\\
\end{eqnarray}
where $V({\bm q},\omega) = v_{\bm q}/\epsilon({\bm q},\omega)$, $v_{\bm q} = 2\pi e^2/q$ is the bare Coulomb interaction and $\epsilon({\bm q},\omega) = 1 - v_{\bm q}\chi_{nn}({\bm q},\omega)$ is the dielectric function. In Eq.~(\ref{eq:Wee_def}), we have defined  the overlap between initial and final states as ${\cal D}({{\bm k},\lambda}; {\bm k}',\lambda') = \big| \langle {\bm k}, \lambda|{\bm k}',\lambda'\rangle \big|^2$, where $|{\bm k},\lambda\rangle$ is an eigenstate of the bare Hamiltonian. For future purposes we define $n_{\rm F}(x) = [ \exp(x) + 1]^{-1}$ and $n_{\rm B}(x) = [ \exp(x) - 1]^{-1}$ as the Fermi and Bose distribution, respectively. 

We solve Eq.~(\ref{eq:boltzmann_def}) in the steady state and to linear order in the electric field by employing the following Ansatz:
\begin{eqnarray} \label{eq:ansatz}
f_{{\bm k},\lambda}  = f^{(0)}(\varepsilon_{{\bm k},\lambda}) -e \tau \frac{{\bm E}\cdot{\bm v}_{{\bm k},\lambda}}{k_{\rm B} T} f^{(0)}(\varepsilon_{{\bm k},\lambda}) \big[ 1 - f^{(0)}(\varepsilon_{{\bm k},\lambda}) \big]
~,
\end{eqnarray}
where $f^{(0)}(\varepsilon_{{\bm k},\lambda})$ is the equilibrium Fermi-Dirac distribution function at the temperature $T$.
Plugging the Ansatz~(\ref{eq:ansatz}) into Eq.~(\ref{eq:boltzmann_def}) and linearizing with respect ${\bm E}$ to we get
\begin{eqnarray} \label{eq:I_ee_2}
\left(-\frac{\partial f^{(0)}(\varepsilon_{{\bm k},\lambda})}{\partial \varepsilon_{{\bm k},\lambda}}\right) {\bm v}_{{\bm k},\lambda}\cdot{\bm E} &=& -\frac{2\pi \tau}{\hbar {\cal A}^3 k_{\rm B} T} \sum_{\substack{{\bm k}_2,\lambda_2}} \sum_{\substack{{\bm k}_3,\lambda_3\\{\bm k}_4,\lambda_4}} |V_{\rm ee}({{\bm k_1}-{\bm k_3}},\varepsilon_{{\bm k}_1,\lambda_1} - \varepsilon_{{\bm k}_3,\lambda_3})|^2 
{\cal D}({{\bm k}_1,\lambda_1}; {\bm k}_3,\lambda_3) {\cal D}({{\bm k}_2,\lambda_2}; {\bm k}_4,\lambda_4) 
\nonumber\\
&\times&
\delta({\bm k}_1+{\bm k}_2-{\bm k}_3-{\bm k}_4) \delta(\varepsilon_{{\bm k}_1,\lambda_1}+\varepsilon_{{\bm k}_2,\lambda_2}-\varepsilon_{{\bm k}_3,\lambda_3}-\varepsilon_{{\bm k}_4,\lambda_4})
\nonumber\\
&\times&
f^{(0)}(\varepsilon_{{\bm k}_1,\lambda_1}) f^{(0)}(\varepsilon_{{\bm k}_2,\lambda_2}) \big[1-f^{(0)}(\varepsilon_{{\bm k}_3,\lambda_3})\big]\big[1-f^{(0)}(\varepsilon_{{\bm k}_4,\lambda_4})\big]
\nonumber\\
&\times&
( {\bm v}_{{\bm k}_1,\lambda_1} +  {\bm v}_{{\bm k}_2,\lambda_2} -  {\bm v}_{{\bm k}_3,\lambda_3} -  {\bm v}_{{\bm k}_4,\lambda_4})\cdot{\bm E}
~.
\end{eqnarray}
To determine the transport time $\tau$, we multiply Eq.~(\ref{eq:I_ee_2}) by ${\bm v}_{{\bm k},\lambda}$ and sum over all ${\bm k}$ and $\lambda$. We obtain
$\tau = D I^{-1}$, where the Drude weight is
\begin{eqnarray}
D = \frac{1}{2 {\cal A}}\sum_{{\bm k},\lambda} \left(- \frac{\partial f^{(0)}(\varepsilon_{{\bm k},\lambda})}{\partial \varepsilon_{{\bm k},\lambda}}\right) |{\bm v}_{{\bm k},\lambda}|^2
~,
\end{eqnarray}
while
\begin{eqnarray} \label{eq:I_ee_2_proj_curr_3}
I &=& 
-\frac{\pi}{16\hbar k_{\rm B} T {\cal A}^3} \sum_{{\bm q}} \int_{-\infty}^{\infty} d\omega \frac{|V_{\rm ee}({\bm q},\omega)|^2}{\displaystyle \sinh^2\left(\frac{\omega}{2k_{\rm B}T}\right)}
\sum_{{\bm k}, {\bm k}'} \sum_{\lambda,\lambda'} \sum_{\eta,\eta'}
{\cal D}({{\bm k},\lambda}; {\bm k}-{\bm q},\lambda') {\cal D}({{\bm k}',\eta}; {\bm k}'+{\bm q},\eta') 
\nonumber\\
&\times&
\big[f^{(0)}(\varepsilon_{{\bm k},\lambda}) - f^{(0)}(\varepsilon_{{\bm k},\lambda}-\omega)\big] \big[f^{(0)}(\varepsilon_{{\bm k}',\eta}) - f^{(0)}(\varepsilon_{{\bm k}',\eta}+\omega)\big]
\nonumber\\
&\times&
| {\bm v}_{{\bm k},\lambda} + {\bm v}_{{\bm k}',\eta}  - {\bm v}_{{\bm k}-{\bm q},\lambda'} - {\bm v}_{{\bm k}'+{\bm q},\eta'} |^2
\delta(\varepsilon_{{\bm k},\lambda} - \varepsilon_{{\bm k}-{\bm q},\lambda'}-\omega)
\delta(\varepsilon_{{\bm k}',\eta}-\varepsilon_{{\bm k}'+{\bm q},\eta'} +\omega)
~.
\end{eqnarray}
The resistivity is therefore
\be
\rho = \frac{I}{e^2 D^{2}}
~.
\ee

We now specialize this result to the case of twisted bilayer graphene. The two layers are kept at different potentials. Therefore, the two Dirac crossings at the points $K$ and $K'$ of the mini-Brillouin zone are shifted in energy in opposite directions. As such, one of the valleys is hole-doped, while the other is electron doped. When the temperature $T$ is much smaller than the Fermi energy $\varepsilon_{\rm F}$ (which we assume to be equal in modulus for the two valleys), we find
\begin{eqnarray}
D = \frac{N_{\rm f} v_{\rm F}^2}{2} \nu(\varepsilon_{\rm F})
~,
\end{eqnarray}
where $\nu(\varepsilon) = \varepsilon/(2\pi\hbar^2 v_{\rm F}^2)$ is the density of states of a single Dirac cone, and $N_{\rm f} = 8$ is the total number of (spin, valley and layer) fermion flavors. As Eq.~(\ref{eq:I_ee_2_proj_curr_3}), in the low-temperature limit the function $1/\sinh^2[\omega/(2k_{\rm B}T)]$ strongly suppresses contributions at large $\omega$. Expanding the integrand in the limit of $\omega\to 0$ to the leading order, we get
\begin{eqnarray} \label{eq:I_ee_manip_1}
I &\simeq& 
\frac{\pi}{16\hbar k_{\rm B} T} \int \frac{d^2{\bm q}}{(2\pi)^2} \int_{-\infty}^{\infty} d\omega \frac{\omega^2 |V_{\rm ee}({\bm q},0)|^2 }{\displaystyle \sinh^2\left(\frac{\omega}{2k_{\rm B}T}\right)}
\int \frac{d^2{\bm k}}{(2\pi)^2} \int \frac{d^2{\bm k}'}{(2\pi)^2} \sum_{\lambda,\lambda'} \sum_{\eta,\eta'}
{\cal D}({{\bm k},\lambda}; {\bm k}-{\bm q},\lambda') {\cal D}({{\bm k}',\eta}; {\bm k}'+{\bm q},\eta') 
\nonumber\\
&\times&
\delta(\varepsilon_{{\bm k},\lambda} - \varepsilon_{{\bm k}-{\bm q},\lambda'})
\delta(\varepsilon_{{\bm k}',\eta}-\varepsilon_{{\bm k}'+{\bm q},\eta'})
\left(-\frac{\partial f^{(0)}(\varepsilon_{{\bm k},\lambda})}{\partial \varepsilon_{{\bm k},\lambda}}\right) \left(-\frac{\partial f^{(0)}(\varepsilon_{{\bm k}',\eta})}{\partial \varepsilon_{{\bm k}',\eta}}\right) 
|{\bm v}_{{\bm k},\lambda} + {\bm v}_{{\bm k}',\eta} - {\bm v}_{{\bm k}-{\bm q},\lambda'} - {\bm v}_{{\bm k}'+{\bm q},\eta'} |^2
~.
\nn
\end{eqnarray}
We now observe that, in the limit of low temperature,
\be
\left(-\frac{\partial f^{(0)}(\varepsilon_{{\bm k},\lambda})}{\partial \varepsilon_{{\bm k},\lambda}}\right) \to \delta(\varepsilon_{{\bm k},\lambda} - \varepsilon_{\rm F})
~.
\ee
Because of the four $\delta$-functions in the integrand of Eq.~(\ref{eq:I_ee_manip_1}), all initial and final states are bound to the Fermi surface. Hence, the matrix element on its last line can be written as
\be \label{eq:matr_el}
{\cal M} \equiv |{\bm v}_{{\bm k},\lambda} + {\bm v}_{{\bm k}',\eta} - {\bm v}_{{\bm k}-{\bm q},\lambda'} - {\bm v}_{{\bm k}'+{\bm q},\eta'} |^2
\simeq
\frac{v_{\rm F}^2 q^2}{k_{\rm F}^2}  |s_\lambda - s_{\eta}|^2
~.
\ee
Here, $s_\lambda=\pm$ accounts for the direction of velocity with respect to momentum in a given mini-valley of the Brillouin zone. To obtain this result, we have neglected the possibility of inter-minivalley transitions. Hence, the particles labelled by $\lambda$ and $\lambda'$ ($\eta$ and $\eta'$) belong to the same mini-valley in the Brillouin zone. This means that they share the same band, valley, and layer index. 

From Eq.~(\ref{eq:matr_el}), it is clear that if all minivalleys are populated with the same type of carriers, the collision integral of electron-electron interactions vanishes. Since, in the present case, populations are unequal, {\it i.e.} $s_\lambda\neq s_\eta$ for some choices of $\lambda$ and $\eta$. To be specific, for each of the $N_{\rm f}$ choices of $\lambda$, there are $N_{\rm f}/2$ possible choices for $\eta$ such that $s_\lambda\neq s_\eta$. In these cases,
${\cal M} = 4v_{\rm F}^2 q^2/k_{\rm F}^2$. Therefore,
\begin{eqnarray} \label{eq:I_ee_manip_2}
I &\simeq& 
\frac{\pi N_{\rm f}^2}{8\hbar k_{\rm B} T} \frac{v_{\rm F}^2}{k_{\rm F}^2} 
\int \frac{d^2{\bm q}}{(2\pi)^2} \int_{-\infty}^{\infty} d\omega \frac{\omega^2 |V_{\rm ee}({\bm q},0)|^2 q^2}{\displaystyle \sinh^2\left(\frac{\omega}{2k_{\rm B}T}\right)}
\Gamma^2(q)
~,
\end{eqnarray}
where
\be
\Gamma(q) &=&
\int \frac{d^2{\bm k}}{(2\pi)^2} \frac{1+\cos(\varphi_{\bm k}-\varphi_{{\bm k}+{\bm q}})}{2} \delta(\varepsilon_{{\bm k}}-\varepsilon_{\rm F})\delta(\varepsilon_{{\bm k}+{\bm q}} - \varepsilon_{\rm F})
\nn
&\simeq&
\Theta(2k_{\rm F}-q) \nu(\varepsilon_{\rm F}) \frac{1}{\pi \hbar v_{\rm F} q}\sqrt{1- \frac{q^2}{4k_{\rm F}^2}} 
~.
\ee
After some lengthy algebra we find 
\be
I &\simeq&
\frac{2\pi^2 N_{\rm f}^2}{3} \left(\frac{k_{\rm B} T}{\varepsilon_{\rm F}}\right)^2 \hbar v_{\rm F}^4 \nu^2(\varepsilon_{\rm F}) \alpha_{\rm ee}^2
\frac{3(N_{\rm F}\alpha_{\rm ee} - 1)+ \big[4 - 3 (N_{\rm F}\alpha_{\rm ee})^2\big] {\rm arccoth}(1+N_{\rm F}\alpha_{\rm ee})}{2}
,
\ee
and therefore the scattering rate and resistivity read
\be
\frac{1}{\tau} = \frac{\varepsilon_{\rm F}}{\hbar}
\frac{2\pi N_{\rm f}}{3} \left(\frac{k_{\rm B} T}{\varepsilon_{\rm F}}\right)^2 \alpha_{\rm ee}^2
\frac{3(N_{\rm F}\alpha_{\rm ee} - 1)+ \big[4 - 3 (N_{\rm F}\alpha_{\rm ee})^2\big] {\rm arccoth}(1+N_{\rm F}\alpha_{\rm ee})}{2}
,
\ee
and
\be
\rho_{\rm el} &=& 
\frac{h}{e^2} 
\frac{4\pi}{3} \left(\frac{k_{\rm B} T}{\varepsilon_{\rm F}}\right)^2 \alpha_{\rm ee}^2
\frac{3(N_{\rm F}\alpha_{\rm ee} - 1)+ \big[4 - 3 (N_{\rm F}\alpha_{\rm ee})^2\big] {\rm arccoth}(1+N_{\rm F}\alpha_{\rm ee})}{2}
,
\ee
respectively.
\end{widetext}


\end{document}